\documentclass[aps,pre,twocolumn,showpacs,amssymb]{revtex4}
\usepackage[dvips]{graphicx}
\usepackage{dcolumn}
\usepackage{bm}
\def\gs{\gtrsim}
\def\ls{\lesssim}
\def\be{\begin{equation}}
\def\en{\end{equation}}    
\def\gs{\gtrsim}
\def\ls{\lesssim}
\newcommand{\bi}[1]{\mbox{\boldmath$#1$}}

\def\p{\partial}
\def\bea{\begin{eqnarray}}
\def\ena{\end{eqnarray}}

\renewcommand{\theequation}{\arabic{section}.\arabic{equation}}

\begin{document}
\draft
\bibliographystyle{prsty}
\title{Pre-dewetting transition on a hydrophobic wall:
Statics and dynamics }
\author{Ryohei  Teshigawara  and Akira Onuki}
\address{Department of Physics, Kyoto University, Kyoto 606-8502}
\date{\today}

\begin{abstract} 
For one-component fluids, we predict 
  a   pre-dewetting  phase transition between  a thin and thick low-density  
 layer in liquid on a wall   repelling   the fluid. 
This is the case of a  hydrophobic wall for water.  
A pre-dewetting line  starts   
from the coexistence curve and ends   
 at a surface critical point in the phase diagram. 
 We  calculate this line   
numerically using  the van der Waals model and analytically 
using the free energy expansion up to the quartic order. 
We  also examine   the pre-dewetting dynamics 
 of a  layer created on 
a hydrophobic spot on a heterogeneous wall. It is from 
a thin to thick layer  during  decompression 
and  from a thick to thin  layer 
 during compression. Upon the  transition, 
 a liquid region above the film 
is cooled for decompression and 
heated for compression 
 due to latent heat convection and a 
 small pressure pulse is emitted from the  film into the liquid. 
\end{abstract}


\maketitle

\pagestyle{empty}

\section{Introduction}
\setcounter{equation}{0}

Extensive efforts 
have been made on the  
wetting transitions  for various liquids  and walls  
both theoretically and experimentally 
 \cite{Cahn,PG,Bonnreview}. 
As is well-known, when a liquid droplet is 
placed on   a wall in gas, 
the three-phase contact  angle changes from a finite 
value (partial wetting) 
to zero (complete wetting) 
 at a wetting transition temperature $T_{\rm w}$ 
 on the coexistence curve. 
Furthermore, there is a phase transition of adsorption 
between a thin and thick liquid layer 
across a prewetting line outside 
the coexistence curve \cite{Cahn,Ebner,Bonnreview,Evans}. 
For  one-component fluids,  
the line  starts from the wetting 
transition point  $T= T_{\rm w}$ 
on the coexistence curve 
and ends at a surface critical point  at  $T= T_c^{\rm pw}$  
in the $T$-$n$ or $T$-$p$ plane, 
where $n$ and $p$ are the density and the pressure 
in  the surrounding gas region.

On the other hand, many authors have been interested in the 
structural change in the hydrogen bonding network 
formed by the water molecules in the vicinity of a hydrophobic 
surface \cite{Is,Sti,Pratt,Ro,Berne,Chandler1999,Ash,Koishi,Netz,Chandler}. 
We also mention many observations of surface bubbles 
or films on mesoscopic scales (with 10-100 nm thickness)  
on hydrophobic walls  in water 
\cite{Higashi,Zhang,Attard,Loshe}. 
Here  the attractive interaction  
among the water molecules  arise from 
the hydrogen bonding. 
As a result, there can even be 
a gas region 
in contact with a hydrophobic surface   in liquid water 
at room temperature and 
at the atmospheric pressure ($\sim 1$atom). 
To support this behavior, 
 the solvation free energy $\Delta G_{\rm sol}$  
of a hydrophobic particle with radius $R$ 
in water is  nearly given by    
$4\pi R^2\sigma$ per particle 
for relatively large $R( \gs 1$nm)  \cite{Chandler}, 
where $\sigma$ is the  gas-liquid surface tension.

In this paper, we  examine whether or not 
 a low-density 
  film in liquid on a hydrophobic wall 
undergoes   a pre-dewetting 
phase transion between a thin and thick low-density  layer. 
As in the   prewetting case,  
a  pre-dewetting line  should start 
from a point, $T=T_{\rm cx}^{dw}$, 
 on the coexistence curve and  ends at a pre-dewetting 
 critical point. It follows 
  complete dewetting for $T>T_{\rm cx}^{dw}$ 
on the coexistence curve. 
 We  develop a  mean-field 
theory based on the Ginzburg-Landau model 
as in the original paper \cite{Cahn}. 
Our calculations are thus performed 
rather  close to the critical point (at $T\sim 0.9 T_c)$, 
where the film density is not very small 
compared to the ambient liquid density.

Furthermore, 
we are interested in the  dynamics 
of the wetting transition  of volatile liquids,  
where understanding of evaporation 
and condensation at the interface  is still 
inadequate \cite{Bonn,Hardy,Dussan}. For example,   
Koplik {\it et al.} \cite{p3}  performed 
 molecular dynamic simulation  to 
observe evaporation of a droplet and a decrease 
of the contact angle upon heating a wall 
in partial wetting. 
Gu$\acute{\rm e}$na {\it et al.} \cite{Caza} 
performed an  experiment, where  
a weakly volatile droplet 
spread  as  an  involatile droplet 
  in an initial stage 
but   disappeared  after 
a long time due to evaporation 
in complete  wetting. 
In a near-critical one-component fluid, 
{Hegseth} {\it et al.}  \cite{Hegseth} 
observed that  a bubble was 
attracted to a heated wall even when 
it was completely wetted
by liquid in equilibrium.  
To study such problems, we have recently 
developed a phase-field model 
for compressible fluids 
with inhomogeneous 
temperature, called the  dynamic van der Waals model   
\cite{Onuki}. (See a review on various phase-field theories of 
 fluids \cite{phase1}.)  
 In our framework, 
 we  may  describe the   
 gas-liquid transition  and  
convective latent heat transport 
 without  assuming  any   evaporation formula. 
We then numerically investigated 
 evaporation of a liquid  droplet on 
 a heated substrate 
 \cite{Teshi1} and spreading of a liquid 
 film on a cooled or warmed substrate for 
a one-component fluid  \cite{Teshi2}.  
The lattice Boltzmann method has been 
applied to two-phase fluids  also   
\cite{Y,Pooly,In}. 
However, this method  has not yet been fully  developed   
 to describe evaporation and condensation.

This  paper also presents  simulation results on 
the film dynamics using  the  
numerical method in our previous studies  
\cite{Teshi1,Teshi2}.  We initially start with 
an  equilibrium  thin or thick film  on a wall 
at the bottom 
near the pre-dewetting transition 
and  then cool or heat the temperature at the top.  
(We use "bottom" and "top" 
though we do not assume gravity.)  
Subsequently, the cell is gradually 
decompressed or compressed 
and the pre-dewetting transition is induced in the film.

The organization of this paper is as follows. 
 In  Sec.II, we  will examine the static aspect 
 of the pre-dewetting transition in 
 the Ginzburg-Landau scheme \cite{Cahn,PG}
 in the mean field theory. 
In  Sec.III, we will apply the dynamic van der Waals model 
to investigate the pre-dewetting dynamics 
by cooling and heating the top plate of a cylindrical cell. 
In the appendix, the pre-dewetting 
transition will be examined near the critical point 
by expanding the free energy with respect 
to the density $n$ around the critical density 
$n_c$ up to the quartic order.


\section{Statics}

\subsection{Ginzburg-Landau model }
\setcounter{equation}{0}
We consider a one-component fluid 
in contact with a sold wall  in equilibrium, where 
the number density  $n$ is 
the order parameter.  
Assuming short-ranged forces, we set up  the free energy $F$ 
with the gradient contribution  
as \cite{Onukibook,vander} 
\be 
{F} 
=\int d{\bi r}\bigg[{f(n,T)}  
+ \frac{1}{2} M|\nabla n|^2 \bigg] 
+ \int da \gamma n ,
\en 
where the integral is in the fluid container 
in the first term and on the wall 
surface in the second term ($\int da$ being the 
surface integral). 
As a function of $n$ and $T$, 
 $f= f(n,T)$ is 
the Helmholtz free energy density. In our numerical 
calculation, we will use the simple 
van der Waals form \cite{Onukibook}, 
\be 
f= k_BT n \bigg[\ln [{n\lambda_{\rm th}^3}/({1-v_0n})] 
-1\bigg] -  \epsilon v_0 n^2 , 
\en 
where $v_0$ is the molecular  volume, 
$\epsilon$ is  the magnitude of the 
attractive  pair potential, and 
 $\lambda_{\rm th}= \hbar(2\pi/mk_B T)^{1/2}$ is the 
thermal de Broglie length with $m$ being the molecular mass. 
The coefficient  $M$ of the gradient free energy  
will be assumed to be  
independent of $n$ but proportional to $T$.  
The last term is  the surface free energy 
expressed as the integration  
on the solid surface \cite{Cahn,PG,Bonnreview}, where 
   the second-order term 
 of the form $M\lambda_e^{-1}(n-n_c)^2/2$ (present in the 
original work \cite{Cahn})  is neglected. 
(This is allowable for very large  $|\lambda_e|$ 
 compared to the correlation length  \cite{Binderreview}.) 
In the literature \cite{Bonnreview,Binderreview},  the  so-called 
surface field is given by  $-\gamma$. 
  For water-like fluids,  $\gamma>0$  
for a hydrophobic surface and  $\gamma<0$ 
  for a hydrophilic surface.  For this  surface free energy, 
  the pre-dewetting transition occurs for $\gamma>0$, while 
    the prewetting transition  for $\gamma<0$.

In equilibrium, the space-dependent 
density  $n=n({\bi r})$ in the bulk region 
is determined by 
\be 
\mu - M\nabla^2 n = \mu_0
\en 
where $\mu=\p f/\p n$ is the chemical potential 
and $\mu_0$ is a constant.  The left hand side 
$\mu - M\nabla^2 n$ represents the generalized chemical 
potential including the gradient contribution. 
 On the wall surface 
we have
\be 
M {\bi \nu}\cdot \nabla n= \gamma ,
\en 
where ${\bi \nu}$ is the outward normal unit vector 
(from the wall to the fluid)  on the 
wall surface.

\subsection{Pre-dewetting transition for $n_0>n_c$}

Let us consider an equilibrium 
 one-component fluid in the region $z>0$ 
in contact with a planar substrate with $\gamma>0$ 
placed at $z=0$. In this  one-dimensional geometry,  
all the quantities depend only on $z$. 
The  density $n(z)$  tends 
  to a constant liquid density 
$n_0$ far from the wall. The pressure and the chemical 
potential far from the wall are 
written as $p_0$ and $\mu_0$, respectively, where $p_0
=n_0k_BT/(1-v_0n_0)- v_0\epsilon n_0^2$ for 
the van der Waals model (2.2). 
In equilibrium, we should minimize the grand potential 
$\Omega$ (per unit area) given by 
\be 
\Omega = \int_0^\infty  dz 
[\omega  + \frac{1}{2}M|n'|^2] + \gamma n_s, 
\en 
where $n'=dn/dz$ and $n_s$ is the surface density,  
\be 
n_s=n(0). 
\en 
We introduce   the grand potential  
density as   
\be 
\omega= f -\mu_0 n +p_0 ,
\en 
Use of  the van der Waals form (2.2) gives  
\bea 
\omega
&=& k_BT n\ln\bigg[\frac{n(1-v_0n_0)}{n_0(1-v_0n)} \bigg] 
- k_BT \frac{n-n_0}{1-v_0n_0} \nonumber\\
&&-\epsilon v_0(n-n_0)^2.
\ena 
Far from the wall, $\omega$ behaves as  
$\omega \cong \chi(n_0)  (n-n_0)^2/2 $ and tends to zero. 
We define  
\be 
\chi (n)= \p^2 f/\p n^2 
=1/n^2K_T,
\en  
where  $K_T$
is  the isothermal compressibility.

In the region  $z>0$,  
the density profile $n=n(z)$ is determined by 
$\mu - Mn'' = \mu_0$ from Eq.(2.3),  
where  $n''= d^2 n/dz^2$. We multiply  
this equation  by $n'$ and 
integrate the resultant one with respect to $z$ 
to obtain  \cite{Onukibook}  
\be 
\omega = M|n'|^2/2.
\en 
From Eq.(2.4)  the boundary condition at $z=0$ reads 
\be 
Mn'(0) = \gamma .
\en 
On the basis of   the van der Waals model (2.2), 
we introduce  a microscopic length $\ell$  defined by 
\be 
\ell= (M/2k_BT v_0)^{1/2}.
\en 
Since  $\gamma>0$,  Eq.(2.10) is rewritten as 
\be 
 n'(z)=
 \sqrt{2\omega(n(z))/M}= 
   \ell^{-1}\sqrt{\omega (n(z))/k_BT v_0},
\en 
where $\omega$ is treated as a function of $n$ 
and its $T$ dependence is suppressed. 
Using $\ell$,  we integrate this equation as  
\be 
z=\ell  \int_{n_s}^n  dn/\sqrt{\omega (n)/k_BT v_0}.
\en 
From Eqs.(2.11) and (2.13)  we have  
\be 
\gamma= \sqrt{2M \omega (n_s)}=2\ell 
\sqrt{k_BTv_0\omega(n_s)}.
\en  
Using  Eq.(2.13)  we rewrite Eq.(2.5) as \cite{Cahn}  
\be 
\Omega= \int_{n_s}^{n_0} dn [ \sqrt{2M\omega(n)}- \gamma] 
+\gamma n_0, 
\en 
where the integrand 
 vanishes at the lower bound $n= n_s$ from Eq.(2.15).
\begin{figure}[t]
\begin{center}
\includegraphics[scale=1.0]{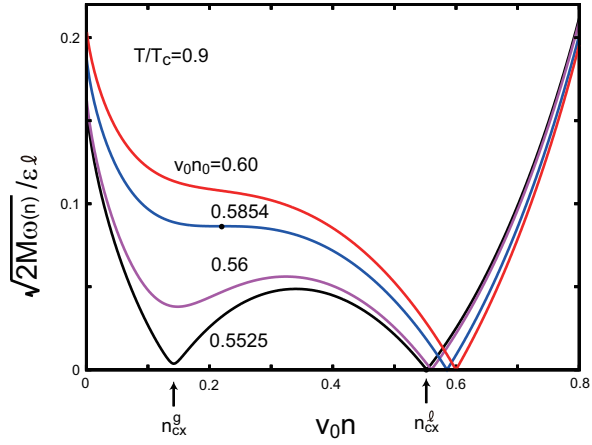}
\caption{\protect  
(Color online) 
$\sqrt{2M \omega(n)}/{\epsilon\ell}$ 
vs $v_0n$ at $T/T_c=0.9$ 
for  $v_0 n_0
=0.5525$, $0.56, 0.58, 0.60$ from below, 
where the smallest one is very close  to 
the coexistence liquid density 
$n_{\rm cx}^\ell=0.5524$. 
Use is made of the van der Waals form (2.8).
}
\end{center}
\end{figure}

\begin{figure}[htbp]
\begin{center}
\includegraphics[scale=1.0]{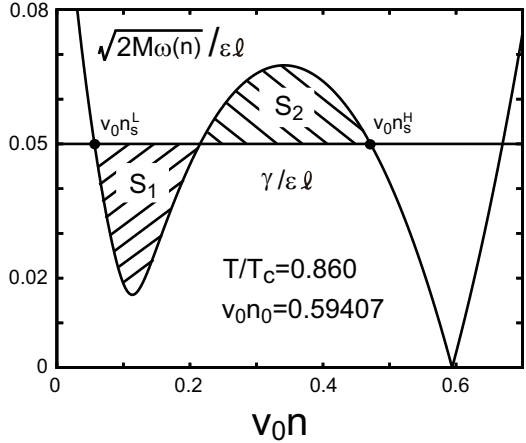}
\caption{\protect
Illustration of a pre-dewetting transition, 
where $\sqrt{2M\omega(n)}/\epsilon\ell$ is plotted 
for $T/T_c=0.860$ and $v_0n_0= 0.59407$.  
At $\gamma/\epsilon \ell=0.05$, 
the areas of the regions $S_1$ and $S_2$ coincide, 
leading to the same value of $\Omega$ in Eq.(2.16) 
at  two surface densities $n_s =n_s^L$ and $n_s^H$.   
}
\end{center}
\end{figure}

 Cahn \cite{Cahn} showed that the thickness 
 of a film of the preferred phase 
on a wall grows logarithmically 
as the surrounding fluid  approaches a state on 
the coexistence curve. In our case,  
the film thickness $D$ increases 
logarithmically as  $n_0$ approaches the  liquid density 
on the coexistence curve $n_{\rm cx}^\ell$, where we may 
take the isothermal path of approach, for example.
To show this, we expand the free energy density $f$  
around the corresponding 
gas density on the coexistence curve 
$n_{\rm cx}^g$ as 
\be 
f\cong  -p_{\rm cx}+ 
\mu_{\rm cx}n+ \chi_g (n-n_{\rm cx}^g)^2/2,
\en 
where $p_{\rm cx}$ and $\mu_{\rm cx}$ are the 
pressure and the chemical potential, 
respectively,  on the coexistence curve. 
When the bulk density $n_0$ is slightly 
larger than  $n_{\rm cx}^\ell$, 
we have  $(p_0-p_{\rm cx})/n_{\rm cx}^\ell
\cong  \mu_0-\mu_{\rm cx} \cong 
 \chi_\ell (n_0-n_{\rm cx}^\ell)$, so that 
\be 
\omega\cong \chi_g (n-n_{\rm cx}^g)^2/2+ \chi_\ell (n_{\rm cx}^\ell
-n_{\rm cx}^g)(n_0-n_{\rm cx}^\ell).
\en 
Substitution of this relation into Eq.(2.14) yields  
\be 
D\cong  \xi_g  \ln \bigg[\frac{2\chi_g (n_{\rm cx}^g-n_s)}{
\chi_\ell (n_0-n_{\rm cx}^\ell)}\bigg],
\en 
where $\chi_g = \chi(n_{\rm cx}^g)$ and  
$\chi_\ell = \chi(n_{\rm cx}^\ell)$ are defined on the coexistence 
curve and $\xi_g=(M/\chi_{\rm cx}^g)^{1/2}$ is the correlation 
length in the gas phase.  
As $n_0- n_{\rm cx}^\ell$ tends to zero, a well-defined 
interface appears.

In Fig.1, we plot $\sqrt{2M\omega(n)}/\epsilon\ell= 
2\sqrt{k_BTv_0\omega(n)}/\epsilon$ 
vs $v_0n$ at $T=0.9T_c$ 
for four bulk densities $n_0$,  
using the van der Waals model  
(2.8). (i) The  smallest  bulk  density is 
$0.5525v_0^{-1}$, which is slightly larger than 
 the coexistence liquid density  
 $n_{\rm cx}^\ell=0.5524v_0^{-1}$. 
 As a result,  $\omega(n)$ nearly  vanishes 
at the   coexistence gas  density 
$n_{\rm cx}^g=0.1419v_0^{-1}$. 
The gas layer thickness $D$ 
is logarithmically dependent on $n_0-n_{\rm cx}^\ell$ as in Eq.(2.19).  
(ii) At $n_0= 0.56v_0^{-1}$, there are a minimum 
and a maximum satisfying 
$\p \omega/\p n=f'-\mu_0=0$. 
(iii) At $n_0= 0.5854 v_0^{-1}$, the two extrema  merge into 
a point $n=n_{\rm sc}$, at which 
$\p^2 \omega/\p n^2= \chi=0$ holds 
and  $n_{\rm sc}$  coincides with the so-called spinodal density 
 on the gas branch  for the linear form 
of the surface free energy. 
(iv) At $n_0= 0.60 v_0^{-1}$,  $\sqrt{\omega(n)}$ 
increases with decreasing $n$ in the range  $n<n_0$. 
 Then Eq.(2.15) yields 
 a unique surface density $n_s$ for any $\gamma>0$.

\begin{figure}[htbp]
\begin{center}
\includegraphics[scale=1.0]{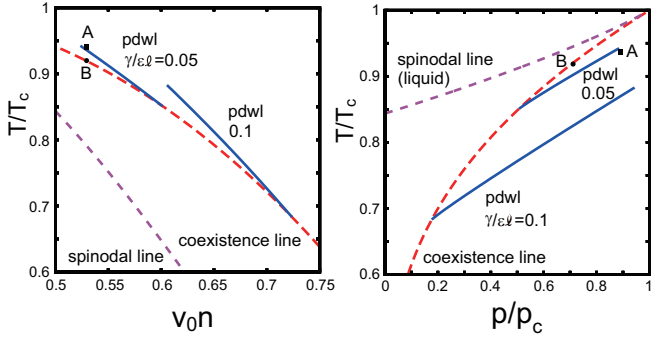}
\caption{\protect 
(Color online) 
Pre-dewetting lines (pdwl) for $\gamma/\epsilon\ell=0.05$ 
and 0.10 in the $T$-$n$ plane (left) 
and in the $T$-$p$ plane (right), where $n$ and $p$ are 
the bulk values  (written as $n_0$ and $p_0$ in the text). 
 Each pdwl   starts from a point on the coexistence curve, 
 where  $T=T_{\rm cx}^{\rm dw}(\gamma)$ and  
 $p=p_{\rm cx}^{\rm dw}(\gamma)$, 
 and ends at a pre-dewetting 
 critical point,  where  $T=T_c^{\rm dw}(\gamma)$ and  
 $p=p_c^{\rm dw}(\gamma)$. See Sec.III for nonequilibrium 
 simulations starting with  points A and B as initial 
 equilibrium states.  
}
\end{center}
\end{figure}

\begin{figure}[htbp]
\begin{center}
\includegraphics[scale=1.0]{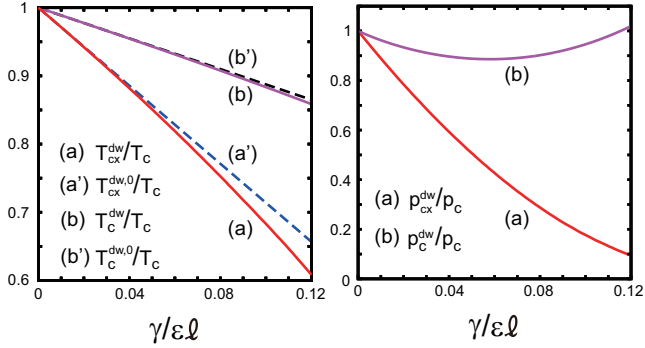}
\caption{\protect 
(Color online) Left: (a) Pre-dewetting temperature 
$T_{\rm cx}^{\rm dw}$ on the coexistence curve 
and  (b) pre-dewetting critical temperature 
$T_c^{\rm dw}$  as functions of $\gamma/\epsilon\ell$. 
Their linear approximations 
are (a') $T_{\rm cx}^{\rm dw,0}$ and (b') $T_c^{\rm dw,0}$ 
(see Eqs.(2.22) and (2.23)). 
Right: (a) Pre-dewetting pressure  
$p_{\rm cx}^{\rm dw}$ on the coexistence curve 
and  (b) pre-dewetting critical pressure 
$p_c^{\rm dw}$  as functions of $\gamma/\epsilon\ell$. 
}
\end{center}
\end{figure}

\begin{figure}[htbp]
\begin{center}
\includegraphics[scale=1.0]{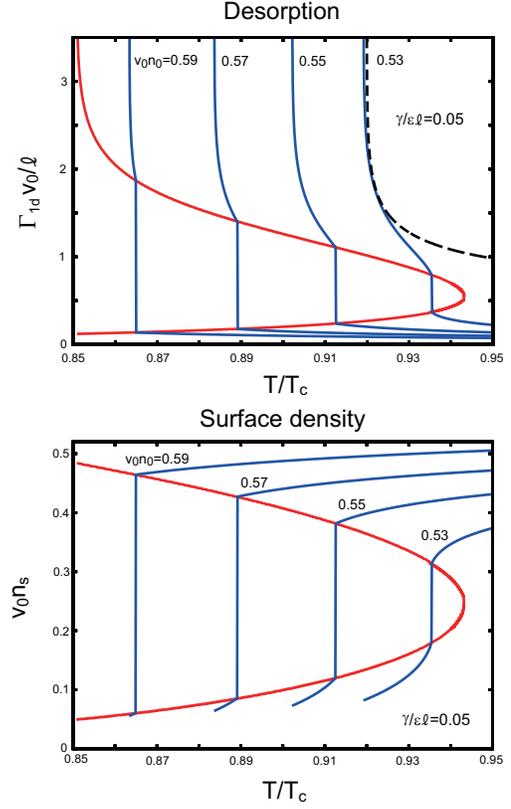}
\caption{\protect 
(Color online) Normalized desorption $v_0\ell^{-1}\Gamma_{1d}$ 
in Eq.(2.25) (top) and normalized surface density $v_0 n_s$ (bottom) 
as functions of $T/T_c$ for $v_0n_0= 0.59,0.57, 0.55,$ and 0.53 
(from left) at $\gamma/\epsilon\ell=0.05$. They change 
 discontinuously  across the pre-dewetting transition. 
Theoretical curve for $v_0\ell^{-1}\Gamma_{1d}$  
from the Landau expansion 
is shown for $v_0n_0=0.53$ (broken line, rightest in top plate).}
\end{center}
\end{figure}

\begin{figure}[htbp]
\begin{center}
\includegraphics[scale=1.0]{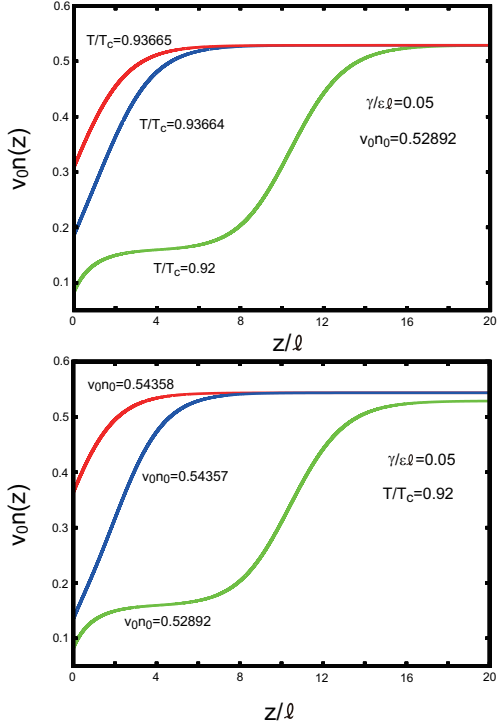}
\caption{\protect
 (Color online) Normalized  density profile $v_0 n (z)$ 
 vs  normalized distance  $z/\ell$ from a wall  at 
  $\gamma/\epsilon\ell=0.05$. Shown are 
  those  for $T/T_c= 0.93665, 0.93664$, and 
0.92 at $v_0n_0= 0.52892$ (top) 
and those  for $v_0n_0= 0.54358, 0.54357$, and 
0.52892 at  $T/T_c= 0.92$ (bottom). The  first two 
curves from above 
are taken just before and after the pre-dewetting transition, 
while the third one close to the coexistence curve.    
}
\end{center}
\end{figure}

As illustrated in Fig.2, a pre-dewetting transition 
appears  when  the curve of $\sqrt{\omega(n)}$ has two extrema 
as in the case of $n_0= 0.56v_0^{-1}$ in Fig.1. 
That is, if the areas $S_1$ and $S_2$ are equal, 
 a first-order   transition occurs 
between two surface densities 
$ n_s^L$ and $n_s^H$.  
For each given  $\gamma>0$, a pre-dewetting 
line starts from a temperature, $T=T_{\rm cx}^{\rm dw}(\gamma)$, 
  on the coexistence curve 
and ends at a pre-dewetting critical temperature, 
 $T=T_c^{\rm dw}(\gamma)$, outside the coexistence curve. 
The wall is completely dewetted by the gas phase 
for $T>T_{\rm cx}^{\rm dw}(\gamma)$ on the coexistence curve.

In Fig.3, two examples of the pre-dewetting line 
are  written for $\gamma/\epsilon\ell=
0.05$ and 0.1 in the $T$-$n$ plane (left) 
and in the $T$-$p$ plane (right), 
where $n$ and $p$ denote  those in the bulk ($n_0$ and $p_0$).   
In this paper,  the coefficient $M$ 
is independent of $n$ and is proportional to $T$ as  
\be 
M= T C .
\en 
Then $C$ and hence 
$\ell= (C/2v_0k_B)^{1/2}$ are independent of $n$ and $T$.
However, essentially the same results were obtained 
even if  $M$ is independent of $T$ (not shown 
in this paper).  We give $T_{\rm cx}^{\rm dw}$ and $ T_{c}^{\rm dw}$ 
and the corresponding densities 
$n_{\rm cx}^{\rm dw}$ and $ n_{c}^{\rm dw}$.  
For $ \gamma/\epsilon\ell=0.05$, we have 
$(T_{\rm cx}^{\rm dw}/T_c, v_0n_{\rm cx}^{\rm dw})
= (0.851, 0.601)$ 
and 
$(T_{c}^{\rm dw}/Tc, v_0n_{c}^{\rm dw})= 
(0.943, 0.524)$. 
For $ \gamma/\epsilon\ell=0.1$, we have 
$(T_{\rm cx}^{\rm dw}/T_c, v_0n_{\rm cx}^{\rm dw})
=(0.683, 0.724)$ 
and 
$(T_{c}^{\rm dw}/T_c, v_0n_{c}^{\rm dw})= 
(0.883, 0.606)$. 

In Fig.4, the left (right) panel displays 
the pre-dewetting transition temperature (pressure) 
$T_{\rm cx}^{\rm dw}(\gamma)$ 
($p_{\rm cx}^{\rm dw}(\gamma)$) on the coexistence curve 
and the pre-dewetting critical  temperature 
(pressure) $T_c^{\rm dw}(\gamma)$ 
($p_c^{\rm dw}(\gamma)$) as functions of $\gamma/\epsilon\ell$.  
For each $\gamma$, the pre-dewetting line is 
in the range $T_{\rm cx}^{\rm dw}(\gamma)<T<T_c^{\rm dw}(\gamma)$. 
If $\gamma/\epsilon\ell\ll 1$, 
these temperatures are both close to $T_c$ 
and are expanded with respect to $\gamma$ as  
\bea 
&& T_{\rm cx}^{\rm dw}(\gamma)= 
T_c (1- A_{\rm cx}\gamma/\epsilon\ell+\cdots), \\  
&& T_c^{\rm dw}(\gamma)= 
T_c (1- A_c\gamma/\epsilon\ell+\cdots).  
\ena 
On the basis of the van der Waals model (2.2), 
the coefficients $A_{\rm cx}$ and $A_{c}$ are 
calculated in the appendix  as 
\be 
A_{\rm cx}= \frac{9}{8} \sqrt{2\sqrt{3}+3}, \quad 
A_{c}= \frac{9}{8}.  
\en 
We also plot the linear approximations 
 $T_{\rm cx}^{\rm dw,0}(\gamma)= 
 T_c (1- A_{\rm cx}\gamma/\epsilon\ell)$ 
 and $T_c^{\rm dw,0}(\gamma)= 
T_c (1- A_c\gamma/\epsilon\ell)$, which are indeed 
in good agreement with the numerical curves for small 
$\gamma\ll \epsilon\ell$.
For the usual  prewetting transition 
in the case  $\gamma<0$,   the counterparts of 
$T_{\rm cx}^{\rm dw}$ and $T_{c}^{\rm dw}$ 
are the wetting temperature $T_{\rm w}$ 
on the coexistence curve 
and the prewetting  critical temperature 
 $T_c^{\rm pw}$. 
 As will be discussed in the appendix, 
 they  are expanded as in Eqs. (2.21) and (2.22) 
 if $\gamma$ is replaced by $|\gamma|$.

In Fig.5, we show the desorption 
$\Gamma_{1d}$ per unit area in the top plate 
 and the surface density $ n_s$ in the bottom plate  
as functions of $T/T_c$ for $v_0n_0= 0.59,0.57, 0.55,$ and 0.53 
 at $\gamma/\epsilon\ell=0.05$.  These lines are 
 outside the coexistence curve in the $T$-$n$ plane.   
They change  discontinuously  across the 
pre-dewetting transition 
 in the range $T_{\rm cx}^{\rm dw}<T<T_{c}^{\rm dw}$.
The discontinuities vanish as $T \to T_{c}^{\rm dw}$.
We calculate  the desorption of the fluid by
\be
\Gamma_{1d} =\int_0^{\infty} dz[n_0-n(r, z)]. 
\en
On  each line, as $T \to T_{\rm cx}^{\rm dw}$,  
$\Gamma_{1d}$ grows logarithmically as $ (n_0-n_{cx}^g)D$ with 
$D$ being  given by Eq.(2.20), while $n_s$ tends 
a well-defined limit (see the lower panel of Fig.5). The curve for 
$v_0n_0=0.53$ is closest to the pre-dewetting criticality, 
so we also display 
its theoretical approximation from the 
Landau expansion in the appendix.

In Fig.6, we display the density profile $n (z)$ 
at $\gamma/\epsilon\ell=0.05$. 
In the top plate 
we set  $T/T_c= 0.93665, 0.93664$, and 
0.92 at   fixed $n_0= 0.52892v_0^{-1}$, 
while in the bottom plate we set 
$v_0n_0= 0.54358, 0.54357$, and 
0.52892 at  fixed $T = 0.92T_c$. 
In these plates,  the first two curves 
represent the profiles 
just before and after the pre-dewetting transition, 
while the third one is obtained 
close to the coexistence curve with a well-defined 
thick gas layer. 

\subsection{Rough estimates of $C$ and $\gamma$ for water}

In our continuum theory, the constant $C$ in Eq.(2.21) 
and the length $\ell$ in Eq.(2.12) 
remain arbitrary.  For each fluid, we may roughly 
estimate their sizes with input of  
the experimental values of 
$T_c$, $n_c$, and the 
surface tension $\sigma$ at some temperature $T$.  
For  example,   we have 
$T_c=647$K and    $n_c= 1.076\times 10^{22}$cm$^{-3}$ for water.   
Using   these $T_c$ and $n_c$ 
in  the van der Waals model calculation, 
we have $\epsilon =28k_BT_c/8=3.0\times 10^{-17}$mJ   
and $v_0= (3n_c)^{-1}=31\times 10^{-24}$cm$^3$, 
from which we  define the van der Waals 
radius by $a_{\rm vdw}= v_0^{1/3}= 3.1 {\rm \AA}$.
For water,  an experimental surface tension is   
$\sigma= 9.3$mJ$/$m$^2$ 
$=0.030 \epsilon/a^2$   at $T=0.92T_c$ \cite{Kislev}. 
On the other hand, from the model (2.1) 
under  Eqs.(2.12) and (2.21), the surface tension 
is numerically calculated as   
$\sigma_{\rm vdw}
=0.0097 \epsilon \ell/v_0$ at $T=0.92T_c$. 
If we equate  these experimental and numerical 
values of the surface tension, we obtain 
$\ell \cong 3.1 a_{\rm vdw}$ and 
$C/2k_B \sim 10 a_{\rm vdw}^5$ \cite{Kitamura}.
In the same manner,  for argon, we have 
$\epsilon= 0.70  \times 10^{-17}$mJ, 
$v_0= 4.11 \times 10^{21}$cm$^3$, and  
$\gamma = 9.3$mJ$/$m$^2= 0.016 \epsilon\ell$ 
 at $T=0.92T_c$ \cite{Buff}, 
leading to   
$C/2k_B \sim 6.8 a_{\rm vdw}^5$.

For water, we estimate the surface field 
$\gamma$ by 
\be 
\gamma  \sim \sigma/n_{\rm liq} ,
\en  
for a typical hydrophobic surface. Here 
$n_{\rm liq}$ and $\sigma$ are 
some  appropriate liquid density and surface tension. 
We use  the the above-mentioned 
 surface tension for water and set 
$n_{\rm liq} \sim n_c$ to obtain 
$\gamma/\epsilon\ell \sim 0.03$.
This estimation  is based on 
 the behavior 
 of the solvation free energy $\Delta G_{\rm sol}$  
of a large hydrophobic particle in water \cite{Chandler}, 
as discussed in Sec.1.

\subsection{Nanobubbles on a hydrophobic spot} 

\begin{figure}[htbp]
\begin{center}
\includegraphics[scale=1.0]{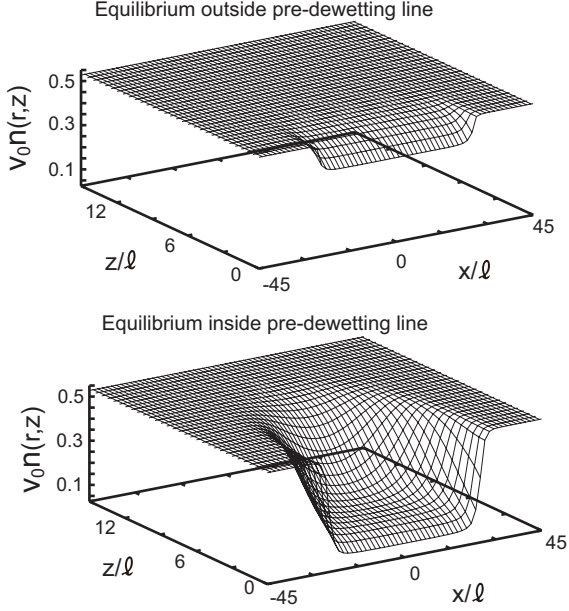}
\caption{\protect
 Equilibrium axisymmetric 
density profile $n(z,r)$ at $y=0$  
on a hydrophobic  spot with radius $25\ell$, 
within which  $\gamma/\epsilon\ell=0.05$ 
and outside which $\gamma=0$. 
Top: $T=0.94T_c$,  $p_0=0.89 p_c$, 
and  $n_0=0.52894v_0^{-1}$ at point A in Fig.3, where 
the fluid is outside the prewetting line 
and the minimum surface 
density is $n_s=0.40587v_0^{-1}$ on the center of the spot. 
 Bottom: $T=0.92T_c$, $p_0=0.71177 p_c$, and 
$n_0=0.52925v_0^{-1}$ at point B in Fig.3,  where 
the fluid is inside the prewetting line 
and the minimum surface 
density is $n_s=0.088433v_0^{-1}$.
}
\end{center}
\end{figure}

Dewetting as well as wetting is very sensitive to 
heterogeneity of the substrate. 
Here we realize   bubbles   in equilibrium on a 
 heterogeneous surface with 
a position-dependent  $\gamma$.

In this paper, we  suppose a circular 
hydrophobic spot with radius 
$25\ell$ on the bottom surface by setting 
\bea 
\gamma(r) &=& 0.05\epsilon\ell \quad (r< 25\ell), \nonumber\\ 
&=&0  ~~~\qquad (r>25\ell),
\ena 
where $r=(x^2+y^2)^{1/2}$ at $z=0$. 
The resultant equilibrium 
density $n(r,z)$ satisfies Eq.(2.3). Its 
 boundary 
condition  at $z=0$ is given by   
$\p n/\p z=0.05\gamma/M= 0.025\epsilon/k_BTv_0\ell$ for $r< 25\ell$  
and $\p n/\p z=0$ for $r>25\ell$.

Slightly above the pre-dewetting line, 
 the  top plate of Fig.7 gives the cross-sectional 
density profile $n(r,z)$ at $y=0$ for the point A in Fig.3, 
where  $T=0.94T_c$, 
$p_0=0.89p_c$, and 
$n_0=0.52894v_0^{-1}$. The minimum density is $n_s=0.40587v_0^{-1}$ 
at the spot center on the surface. 
Below the pre-dewetting line, 
the bottom plate of Fig.7  gives the 
profile at  the point B  in Fig.3, 
where  $T=0.92T_c$, $p_0=0.71177p_c$, and 
$n_0=0.52925v_0^{-1}$.  The minimum density is 
$n_s=0.088433v_0^{-1}$ at the spot center. 
These  profiles are very different. 
To characterize the film size, 
 we introduce  the desorption of the fluid by
\be
\Gamma
=2\pi\int_0^{z_0}dz\int_0^{r_0}dr r[n_0-n(r, z)], 
\en
where the gas film is well within 
the integration region $0<z<z_0$ and $r<r_0$. 
By setting  $z_0=15\ell$ and  $r_0=37.5\ell$, 
we obtain 
$\Gamma=339\ell^3v_0^{-1}$ and 
$3460\ell^3v_0^{-1}$  
for the top 
and bottom plates in Fig.7, respectively.

\setcounter{equation}{0}

\section{Dynamics}

In this section, using  the dynamic van der Waals 
model\cite{Onuki}, we numerically 
 investigate the dynamics of 
a thin gas layer created  on 
the hydrophobic spot in Eq.(2.27). 
We  treat a one-component fluid  without gravity 
in a temperature range 
 $0.92\le T/T_c\le 0.94$, where 
the gas  density is  $25-30\%$ of the 
liquid density (see the bottom curves in Fig.6). Then 
the mean free path in the gas $\ell_{\rm mf}$ 
 is not very long. 
For   very  long $\ell_{\rm mf}$, however, 
numerical analysis based on a  continuum 
phase-field model becomes  very difficult. 
In our simulation, furthermore, 
 relatively small  temperature changes are applied 
 and  heat and mass 
fluxes  passing  through the interface remain  weak.    
 As a result,  $T$ and the generalized 
 chemical potential (the left hand side of Eq.(2.4)) 
 are continuous across the  interface.

In our diffuse interface method, the interface thickness 
needs to be longer than the simulation mesh size $\Delta x$, 
so our system size  cannot be very large. 
In our simulation to follow, 
its  length is  $ 150 \ell$. Nevertheless,  
our cell  contains many particles   about 
$Vn_0 \sim 6 \times 10^6\ell^3/v_0$, where 
$V$ is the volume and 
$n_0$ is the initial liquid density.

\subsection{Hydrodynamic equations with gradient stress}

We set up  the  hydrodynamic equations 
for the mass density $\rho=mn$, the momentum density 
$\rho {\bi v}$, and the 
entropy density $\hat{S}$  \cite{Teshi1,Teshi2}, 
where $m$ is the molecular mass 
and  $\bi v$ is the velocity field.  
Here  $\hat S$  consists of the usual 
entropy density $ns$ and  the negative  gradient entropy as 
\be 
\hat{S}= ns -\frac{1}{2}C|\nabla n|^2 .
\en 
where  $s$ is the entropy per particle 
and the coefficient $C$  in Eq.(2.21) appears here.  
The internal energy density, written as $e$, 
can also contain the gradient contribution, 
but we neglect it for simplicity \cite{Onuki}. 
Then the total  Helmholtz free energy density 
is given by $e-T{\hat S}= 
f+ M|\nabla n|^2/2$ as in the first term as given 
in Eq.(2.1). 
In our scheme, we   use   the entropy equation  
instead of the energy equation 
to achieve the numerical stability in the interface region. 
That is, with our  entropy method, we may remove the so-called 
parasitic flow at the interface, which has been 
encountered by many authors \cite{para}.

 We integrated  
the following hydrodynamic equations without gravity 
for $\rho$, $\rho{\bi v}$, and $\hat S$ 
\cite{Teshi1,Teshi2}:   
\bea
&& 
\frac{\p}{\p t} \rho + \nabla \cdot(\rho{\bi v})=0,\\
&&  
\frac{\p}{\p t}\rho {\bi v}
+\nabla \cdot (\rho  {\bi v}{\bi v}+
 \tensor{\Pi})= \nabla\cdot\tensor{\sigma} , 
\\
&&\frac{\p}{\p t}\hat{S} 
+\nabla\cdot \bigg[
\hat{S}{\bi v}-Cn(\nabla\cdot {\bi v})\nabla n \bigg] 
= \nabla\cdot \frac{\lambda}{T}\nabla T \nonumber\\
&&\hspace{2cm}  + 
({\dot{\epsilon}_v+\dot{\epsilon}_\theta})/{T}, 
\ena
where the terms in the right hand sides are dissipative. 
In Eq.(3.3),  $\tensor{\Pi}=\{\Pi_{ij}\}$ is 
the reversible stress tensor, 
\bea
\Pi_{ij} &=& \bigg [
p  - CT( n\nabla^2 n+\frac{1}{2} |\nabla n|^2)\bigg]\delta_{ij}
\nonumber\\
&&  +CT (\nabla_i n)(\nabla_j n), 
\ena 
where  ${ p}$ is the van der Waals pressure. 
 Hereafter $\nabla_i= \p /\p x_i$ 
with $x_i$ representing $x$, $y$, or  $z$. 
The terms proportional to $C$ arise from the 
gradient entropy, constituting the gradient stress tensor.    
The  $\tensor{\sigma}= \{{\sigma}_{ij}\}$ in the right hand side of Eq.(3.3) 
is the  viscous stress tensor,  
\be 
{\sigma}_{ij}
=\eta(\nabla_i v_j+\nabla_j v_i)  +
(\zeta-2\eta/3) (\nabla \cdot {\bi v})\delta_{ij} ,
\en 
 in terms of the 
shear viscosity $\eta$ and the bulk viscosity 
$\zeta$. In Eq.(3.4), $\lambda$ is 
the thermal conductivity, while 
\be   
\dot{\epsilon}_v= \sum_{ij}\sigma_{ij}\nabla_j v_i,
\quad 
\dot{\epsilon_\theta}= \lambda(\nabla T)^2/T, 
\en  
are the nonnegative entropy production rates 
arising from the viscosities and the thermal conductivity, 
respectively.  On approaching 
equilibrium, $\dot{\epsilon}_v$ 
and $\dot{\epsilon_\theta}$ tend to zero, 
leading to vanishing of the gradients of $\bi v$ 
and $T$.

The (total) energy density in the bulk is defined by  
\be   
e_{\rm T}={e}+ \rho {\bi v}^2/2,  
\en  
which includes   the kinetic energy density. 
The  energy-conservation 
 equation reads \cite{Landau}   
\be 
\frac{\p}{\p t} {e}_{\rm T}=
 -\nabla\cdot\bigg[ e_{\rm T}{\bi v} + 
(\tensor{\Pi}-\tensor{\sigma})\cdot{\bi v} 
-\lambda \nabla T\bigg]. 
\en 
If  Eqs.(3.2) and (3.3) are assumed, 
the entropy equation (3.4) and 
the energy equation (3.9) are obviously equivalent. 
In the text book \cite{Landau}, however,    
the entropy equation  is 
derived from the three fundamental conservation   
equations (3.2), (3.3), and (3.9) 
Now the total fluid 
entropy ${\cal S}_{\rm tot}$ 
and the total fluid energy ${\cal E}_{\rm tot}$ 
are written as   
\be 
{\cal S}_{\rm tot}= \int d{\bi r} \hat{S}, \quad   
{\cal E}_{\rm tot}= \int d{\bi r}e_{\rm T}
+ \int da \gamma n.
\en 
If the surface field $\gamma$ 
is independent of  $T$ as in Eq.(2.26), 
the surface term in Eq.(2.1) is the surface energy 
and there 
is no surface entropy on all the 
boundaries. 
In a  fixed cell, we assume  the no-slip condition 
${\bi v}={\bi 0}$ on its  boundaries. 
Then  the space integrations 
of Eq.(3.4) and (3.9) yield 
the time derivatives of 
 $ {\cal S}_{\rm tot}$ and $ {\cal E}_{\rm tot}$:  
\bea 
&&\hspace{-12mm} 
\frac{d}{dt} {\cal S}_{\rm tot}
= \int d{\bi r}\frac{\dot{\epsilon}_v+\dot{\epsilon}_\theta}{T} 
+\int da \frac{{\bi \nu}\cdot\lambda\nabla T 
 +\gamma \dot{n}}{T},\\
&& \hspace{-12mm}
\frac{d}{dt} {\cal E}_{\rm tot}
= \int da ({{\bi \nu}\cdot\lambda\nabla T 
 +\gamma \dot{n}}), 
\ena 
where $\dot{n}=\p n/\p t$ and $\gamma$ 
can be heterogeneous as in  Eq.(2.26). 
If there is no heat input 
and ${\dot n} =0$ 
on the boundary walls, 
$ {\cal S}_{\rm tot}$  continue to increase 
monotonically until an equilibrium  
state is realized.

 \subsection{Simulation method}

We suppose a cylindrical cell in the axisymmetric geometry, 
where our model fluid is in the region 
$0\leq z\leq H$ and $0\leq r=({x^2+y^2})^{1/2} \leq L$.
  Assuming that  all the  variables   depend 
only on $z,$ $r$ and $t$, we perform 
 simulations on a two-dimensional $300\times 300$ lattice. 
We take   the simulation mesh length $\Delta x$ 
equal to $\Delta x=\ell/2$, where   $\ell$ 
 is   defined  in Eq.(2.12).  Thus our cell  is 
 characterized  by   
\be 
H= 150\ell, \quad 
L=150 \ell, 
\en 

The velocity  $\bi v$ vanishes 
on all the boundaries. 
 The viscosities and the thermal conductivities 
are   proportional to $n$ as   
\be 
\eta= \zeta= \nu_0 mn, \quad \lambda=4k_B\nu_0 n.
\en  
These   coefficients 
 are larger in liquid than  in gas by the density ratio 
$n_\ell/n_g (\sim 5$ in our simulation). 
The  kinematic viscosity 
$\nu_0=\eta/mn$ is a constant. 
We will  measure time in units of the viscous relaxation time,  
\be 
\tau_0=\ell^2/\nu_0= C/2k_Bv_0\nu_0,
\en 
on the scale of $\ell$.  The thermal diffusion constant $D_{\rm th}=
\lambda/C_p$ is of order $\nu_0$, 
where  $C_p$ 
is  the isobaric specific heat per unit volume of 
order $k_Bn$ (not very close to the critical point). 
The time mesh  size $\Delta t$ in integrating 
Eqs.(3.2)-(3.4)  is  $0.02\tau_0$. 
If the dynamic equations are made dimensionless, 
there appears a dimensionless number given by 
$\hat{\sigma}\equiv 
m  \nu_0^2/\epsilon\ell^2$ (written as 
$\sigma$ in Ref.\cite{Onuki}), where $m$ is the molecular 
mass. The transport coefficients 
are  proportional to $\nu_0 \propto \hat{\sigma}^{1/2}$. 
In this paper we set $\sigma=0.06$, for which 
sound waves are well-defined as oscillatory modes  
for wavelengths longer than 
$\ell$ \cite{Onuki}.

As the boundary conditions, 
we assume Eq.(2.4) for the density $n$ 
(even in nonequilibrium) so that 
 $C \p n/\p z=0.05 \epsilon\ell/T$ 
 on the hydrophobic spot at $z=0$ 
 and ${\bi \nu}\cdot \nabla n=0$ 
 on all the other surface regions. 
For the velocity $\bi v$, 
the no-slip  condition ${\bi v}= {\bi 0}$ 
is assumed. The temperature 
 is fixed at $T_0$ at the bottom  $z=0$ 
and at $T_H$ at the top  $z=H$.  
The side wall is thermally insulating 
 as  $\p T/\p r=0$ at $r=L$.

\subsection{Decompression by cooling the upper plate}

We prepared the equilibrium state at $T=0.94T_c$  
in the upper plate of Fig.7 as 
an initial state at $t=0$. We then 
cooled the top temperature $T_H$  from 
  $0.94T_c$ to $0.92 T_c$ fixing the bottom temperature 
  $T_0$  at  $0.94T_c$ for $t>0$. 
After a very long time $(t> H^2/4D_{\rm th}$), 
the fluid tended to a steady 
heat-conducting state. 
\begin{figure}[thbp]
\begin{center}
\includegraphics[scale=1.0]{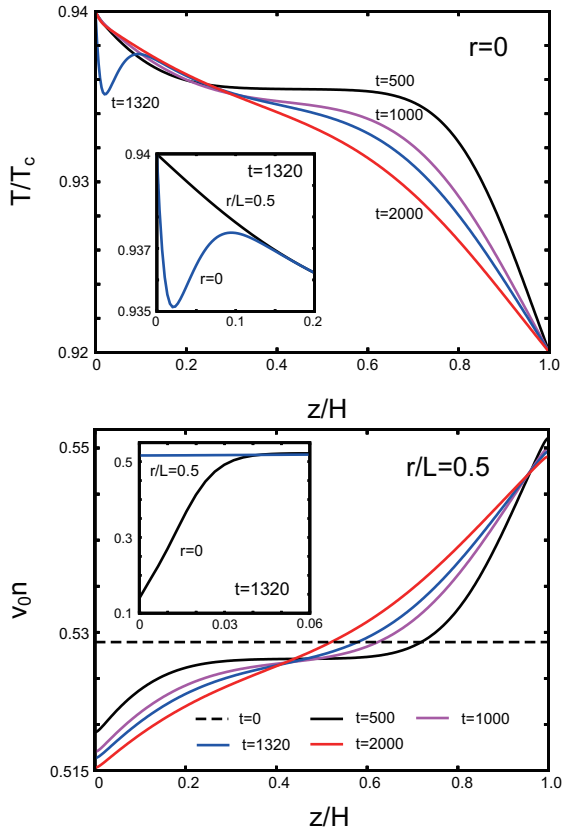}
\caption{\protect 
(Color online) 
Normalized temperature $T(r,z,t)/T_c$  at $r=0$ (top) and 
 density $v_0 n(r,z,t)$  at $r/L=0.5$ (bottom) 
vs $z/H$   after cooling the top from 
$0.94T_c$ to $0.92T_c$. 
At $t =500\tau_0$, thermal diffusion layers appear 
near the top and bottom, causing an  adiabatic change  
in the middle region. For $t \gs 1000\tau_0$, 
the thermal diffusion extends  throughout the cell. 
The density is decreased in the lower part of the cell, 
resulting in a homogeneous pressure decrease.  
Around $t=1320\tau_0$, a cool spot 
appears  above the film during  
 the pre-dewetting transition  from a thin 
to thick film.  In the inset, the profiles of 
$T$ and $n$ at $r=0$ and $L/2$ near the bottom 
are compared.  
}
\end{center}
\end{figure}

\begin{figure}
\begin{center}
\includegraphics[scale=1.0]{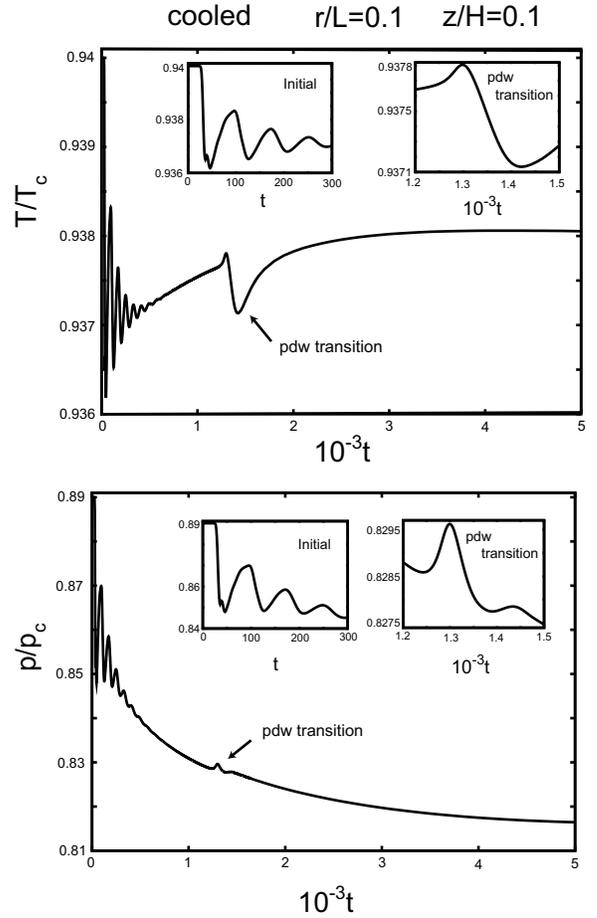}
\caption{\protect 
Normalized temperature $T(r,z,t)/T_c$ (top) and pressure 
$p(r,z,t)/p_c$ (bottom) as functions of  $t$ in units of $\tau_0$ 
in Eq.(3.15) 
at $(z/H, r/L)=(0.1, 0.1)$ (slightly above the gas film) 
 after  cooling  the top. 
Two insets in each panel 
show  initial  oscillatory 
behavior due to sound wave propagation (left) and 
a sudden change   at 
the prewetting transition (right). 
}
\end{center}
\end{figure}

\begin{figure}[htbp]
\begin{center}
\includegraphics[scale=1.0]{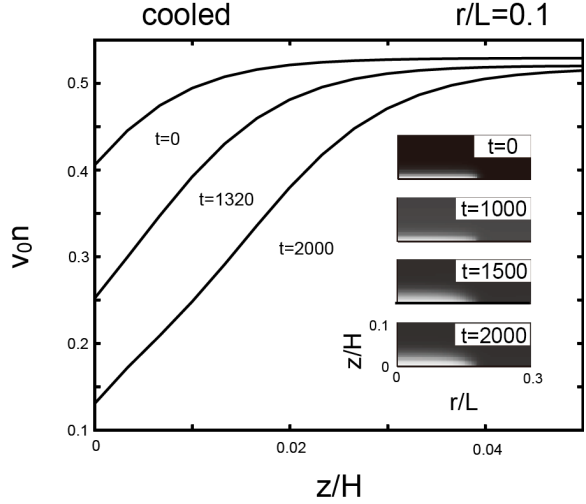}
\caption{\protect
Normalized density  
$v_0 n(r,z,t)$
vs $z/H$  at $r/H=0.1$ 
for  $t/\tau_0=0$, $1320$, and $2000$  from above after 
 cooling  the top.  
At $t=1320\tau_0$, the film thickness 
is increasing abruptly at the pre-dewetting transition, 
while at $t=2000\tau_0$  
it is still increasing slowly (see Fig.11). 
Cross-sectional film profiles are also given 
in the right, where the darkness represents 
the density at four  times.  
}
\end{center}
\end{figure}

\begin{figure}
\begin{center}
\includegraphics[scale=1.0]{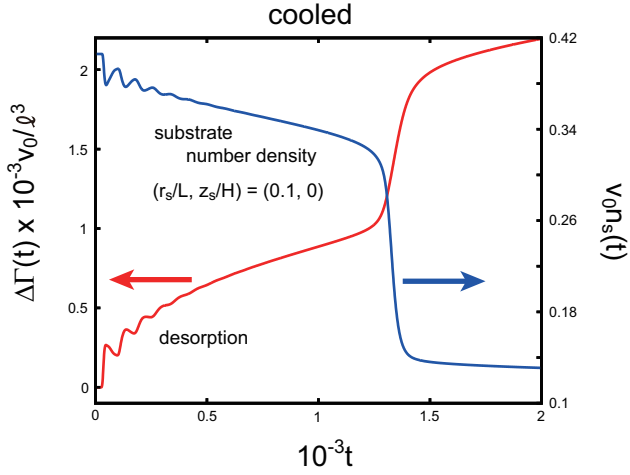}
\caption{\protect
Normalized excess desorption  
$v_0\ell^{-3}\Delta \Gamma(t)$ in Eq.(3.16) 
 and surface density  $v_0n_s(r,t)$ vs $t$   
 at $ r/L= 0.1$  in units of $\tau_0$ 
 after  cooling  the top.    
They exhibit oscillatory behavior in the initial stage 
and an abrupt change 
 at the pre-dewetting transition taking place in a time interval of 
 $250 \tau_0$. 
}
\end{center}
\end{figure}
\begin{figure}[htbp]
\begin{center}
\includegraphics[scale=1.0]{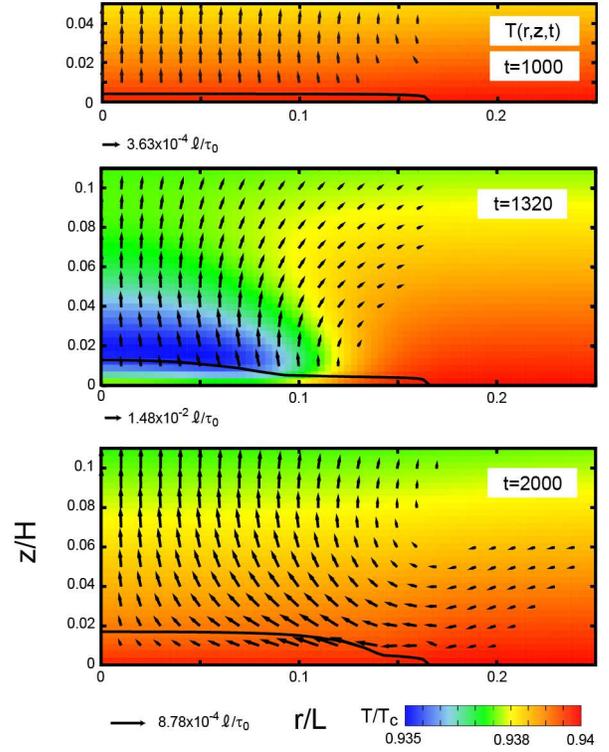}
\caption{\protect 
(Color online)  
Temperature $T(r,z,t)$  in gradation at $t/\tau_0=
1000, 1320$, and $2000$ around the 
hydrophobic spot. On  the block line, 
$|\nabla n|$ is largest, indicating the film location. 
Arrows indicate the velocity  $\bi v$. 
Their  amplitudes are  known  from reference 
arrows  below the middle abd bottom panels. 
The liquid region above the film 
is cooled due to latent heat absorption to the film 
at the pre-dewetting transition. 
The profile at $t=2000\tau_0$ (bottom) is nearly stationary, 
where condensation  is taking from the side 
and  evaporation from the upper surface.  The colors 
represent $T/T_c$ according to the color bar at the bottom.  
}
\end{center}
\end{figure}

In Fig.8, we show the  profiles of the 
 temperature $T(r,z,t)$ at  $r=0$   and the 
density $n(r,z,t)$ at  $r=L/2$  as functions of $z$ 
to  illustrate   how  
 cooling and decompression are realized  
in the cell.  In  the initial stage, the  
 piston effect  is operative, which   
has  been studied  theoretically  \cite{Ferrell,Zappoli} 
and experimentally  \cite{Beysens,Moldover,Miura}.  
In this situation, there appear  
a compressed  thermal diffusion layer 
 at the top and an expanded one  
 at the  bottom  with 
 thickness growing in time as $\ell_{\rm D}(t)= 
 (D_{\rm th}t)^{1/2}$, 
 which produce sound  waves propagating 
 in  the cell and causes an  adiabatic 
 change  outside the 
 diffusion layers. The sound velocity 
 in the present case is about $c= 4\ell/\tau_0$, 
 so the acoustic traversal time over the cell is 
 $H/c \sim 35\tau_0$.    In fact, at $t=500\tau_0$, 
  $T$ and $n$ are flat   in the middle region, where 
 $T$ and $n$ are changed adiabatically.  
 However, in the late stage $t \gs 1000\tau_0$, 
 the layer thickness reaches $H/2$ 
 and  the thermal diffusion  becomes relevant  
 throughout the cell.  In all these processes, the pressure $p$ 
 is kept nearly homogeneous in the cell.

 In our problem,  we should focus on 
 the behavior of $T$ and $p$ around the gas film 
 on the hydrophobic spot. The film   is under 
 gradual decompression,  but    
 nearly at the initial temperature 
 (since the bottom temperature is pinned). 
  As a result, at $t=1320\tau_0$,  
the pre-dewetting transition takes place 
from a thin to thick gas film.  
This is  indicated by formation of 
a cool spot in the liquid above the film in Fig.8. 
It  is caused by  the latent heat 
adsorption to the expanding film 
at the first-order phase transition 
(see Fig.12 in more detail). There is no cooling 
outside the film region $r/L\gs 0.2$.

In Fig.9, we display  the 
 temperature $T(r,z,t)$  and the pressure  $p(r,z,t)$ 
slightly  above the film at a fixed position 
located at  $(z/H, r/L)=(0.1, 0.1)$. 
In the early stage $t\ls 300\tau_0$, 
we can see their oscillatory relaxations  caused by 
sound wave traversals. 
Upon occurrence of the pre-dewetting transition 
at $t \sim 1300 \tau_0$,  
 we can see  a small  drop in $T$ of order $5\times 
 10^{-4}T_c$ 
and a small peak in $p$ of order 
$10^{-3}p_c$ on their curves. 
However, the insets in Fig.9 
reveal  that their behavior 
is somewhat complicated 
on a short time scale of order $50\tau_0$, 
because they change 
on arrivals of a sound wave and a velocity disturbance 
 from the film.   At long  times, 
$T$ tends to a constant about $0.938T_c$,  
but $p$ continues  to decrease slowly.

In Fig.10, 
the  profiles of the density   
$n(r,z,t)$  vs $z$ are given 
 at $r/H=0.1$ at three times 
 together with their cross-sectional  profiles. 
The increase in the film thickness is 
abrupt around  
$t=1320\tau_0$ at the pre-dewetting transition 
and is very slow  at $t=2000\tau_0$. 
  Figure 11  presents  the time evolution 
of the surface density $n_s(r,t)$ 
at $r/L=0.1$   and 
the  excess  desorption $\Delta \Gamma(t)$ defined by 
\be
\Delta\Gamma(t)
=2\pi\int_0^{z_0}\hspace{-2mm}
dz\int_0^{r_0}\hspace{-1mm} 
dr r [(n_0( r,z)-n(t, r,z))],
\en 
where $n_0( r,z)= n(0, r,z)$ is the initial 
density profile and we set  $z_0=0.1H$ and $r_0=0.25L$. 
The initial oscillatory relaxations arise from 
traversals of sound  waves, 
while $n_s$ decreases and 
$\Gamma$ increases abruptly 
at the pre-dewetting transition at $t\sim 1.3\times 10^3\tau_0$. 
We recognize that the pre-dewetting transition from a thin to 
thick film occurs in a time of order $250\tau_0$.

In Fig.12, we show $T(r,z,t)$ in gradation 
at $t/\tau_0=
1000, 1320$, and $2000$ around the 
hydrophobic spot.  Since there is no clear interface here, 
 a gas-liquid boundary is indicated 
by   a line on which 
$|\nabla n|$ is largest in the direction of 
$\nabla n$.  
We also display  the velocity  $\bi v$ by arrows. 
The reference  arrow below each panel represents 
the maximum velocity, being equal to 
0.0363,  1.48, and  0.0878, 
at $t/\tau_0=1000, 1320$, and $2000$, 
respectively,  in units of $10^{-2}\ell/\tau_0$.
It is  enhanced  in  the middle plate 
at $t/\tau_0= 1320$  during 
 the pre-dewetting transition, 
 where we can see an  upward flow with a 
magnitude of $1.5\times 10^{-2}\ell/\tau_0$
in the liquid region above the expanding film. 
The expanding velocity of the film is of the same order.
If we multiply this velocity by the duration time 
$250\tau_0$ of the transition (which is inferred from 
Fig.11), we obtain a  
film thickness of order $D \sim 4\ell$ in accord with 
the density profiles in Fig.10. 
The corresponding Reynolds number around the film 
is about $D|{\bi v}|/\nu_0  \sim 0.1$.  
Remarkably, the region above the film 
is cooled due to latent heat absorption to the film 
by $2\times 10^{-3}T_c$. 
The profile at $t=200\tau_0$ is nearly stationary, 
where the typical velocity is of order  
$5\times 10^{-4}\ell/\tau_0$. 
A balance is attained 
between  condensation  from the side 
and  evaporation from the upper surface, while 
the fluid is at rest far from the film in the presence of 
a constant temperature gardient. 
Finally, we give the heat flux 
$Q_b(r,t)= - \lambda \p T/\p z$  at $z=0$ 
from the wall to the fluid at $r=0$ and $L/2$:  
$(Q_b(0, t), Q_b(L/2,t)= 
(1.39, 1.13)$ at $t=1000$, 
$(5.71, 0.942)$ at $t=1320$, and 
$(0.618, 0.867)$ at $t=2000$ 
in units of $10^{-4} \epsilon\ell / v_0\tau_0$. 
The heat flux is much enhanced  at the center 
during the pre-dewetting transition. 
In the inset of the upper panel of Fig.8, 
the gradient    $|\p T/\p z|$ is very large 
at $r=0$, but it is  much smaller 
at $r=L/2$.

\subsection{Compression by heating 
the upper plate}

We prepared the equilibrium state at $T=0.92T_c$ 
in the lower  plate of Fig.7  (at the point (B) in Fig.3)  
as  an initial state at $t=0$. We then 
heated  the top temperature $T_H$  from 
  $0.92T_c$ to $0.94 T_c$ with the bottom temperature 
  $T_0$  held fixed at  $0.92 T_c$ for $t>0$.

\begin{figure}[htbp]
\begin{center}
\includegraphics[scale=1.0]{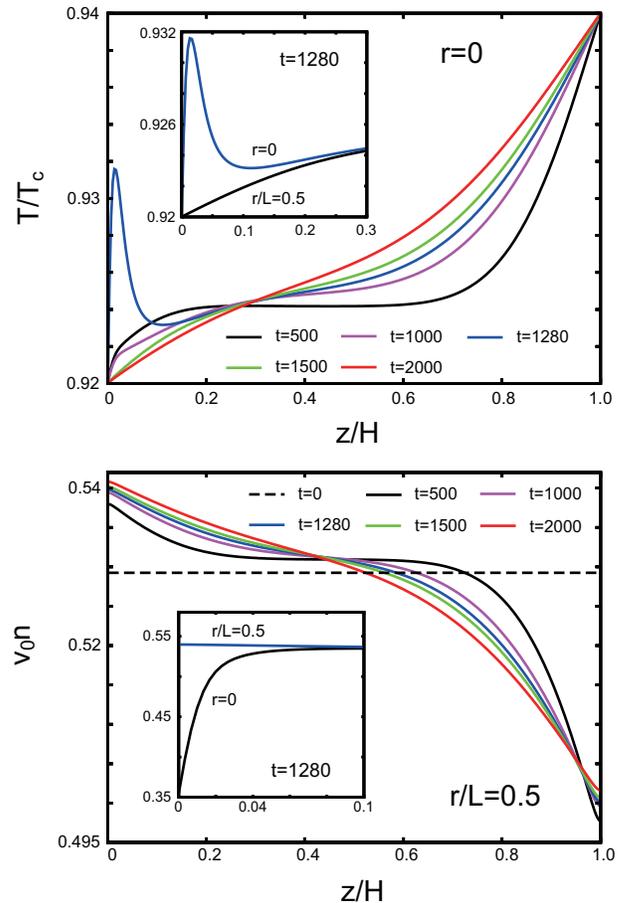}
\caption{\protect 
(Color online) 
 $T(r,z,t)/T_c$  at $r=0$ (top) and 
$v_0 n(r,z,t)$ at $r/L=0.5$ (bottom) vs $z/H$ 
 after heating the top from 
$0.92T_c$ to $0.94T_c$. 
At $t =500\tau_0$, thermal diffusion layers appear 
near the top and bottom. 
For $t \gs 1000\tau_0$, 
the thermal diffusion extends  throughout the cell. 
The density is increased in the lower part of the cell, 
resulting in a homogeneous pressure increase.  
Around $t=1280\tau_0$, a heat  spot 
appears  above the film during  
 the pre-dewetting transition. 
In the inset, the profiles of 
$T$ and $n$ at $r=0$ and $L/2$ 
near the bottom are compared.  
}
\end{center}
\end{figure}

\begin{figure}[htbp]
\begin{center}
\includegraphics[scale=1.0]{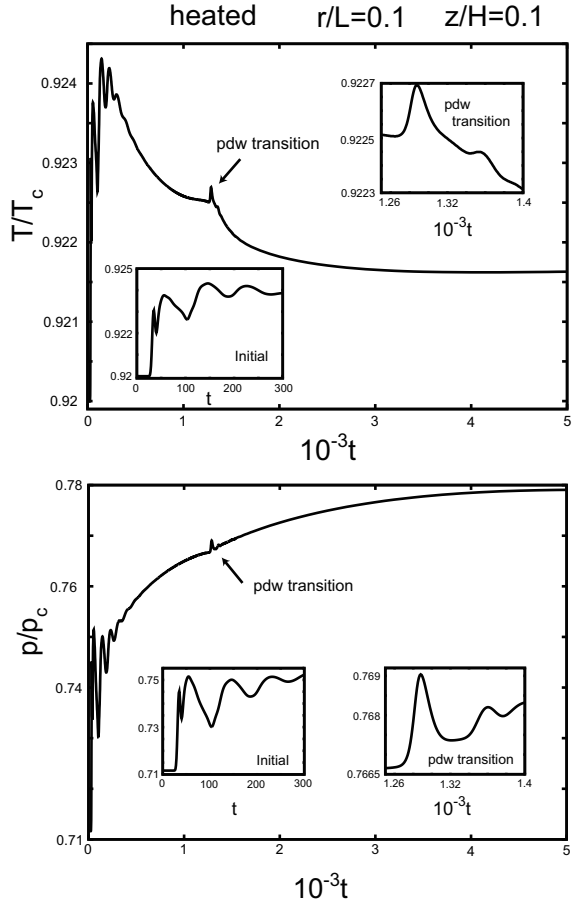}
\caption{\protect 
 $T(r,z,t)/T_c$ (top) and 
$p(r,z,t)/p_c$ (bottom) as functions of  $t$ in units of $\tau_0$ 
at $(z/H, r/L)=(0.1, 0.1)$ (slightly above the film), 
after heating  the top. 
Two insets in each panel 
show  initial  oscillatory 
behavior due to sound waves (left) and 
a sudden change   at 
the pre-dewetting transition from a thick  
to thin film (right).
}
\end{center}
\end{figure}

In Fig.13, the time evolution 
 of  $T(r,z,t)$  at  $r=0$   and  $n(r,z,t)$ at  $r=L/2$ 
  is illustrated 
 as functions of $z$. 
 As in Fig.8,  
  the  piston effect  takes place 
  in  the initial stage, but in the reverse 
 direction. That is, at $t=500\tau_0$, 
 there appear  an expanded thermal diffusion layer 
 at the top and a  compressed one 
 at the  bottom.   
 In the late stage $t \gs 1000\tau_0$, 
  the thermal diffusion  extends over  the cell. 
The pressure $p$ 
is  nearly homogeneous in the cell 
and increases slowly in time 
 above the initial value. 
 Thus the film   is under 
 gradual compression,  but    
 nearly at the initial temperature. 
Occurrence   of 
the pre-dewetting transition 
is indicated by formation of 
a heat spot in the liquid above the film at $t=1280\tau_0$. 
It  is caused by  the latent heat release 
 from the shrinking  film 
(see Fig.17).

Figure 14 displays the temperature  $T(r,z,t)$  and 
the pressure   $p(r,z,t)$ 
at  $(z/H, r/L)=(0.1, 0.1)$ as in Fig.9.   
The adiabatic process takes place 
in the early stage. We 
 find  occurrence of the pre-dewetting transition 
at $t \sim 1280 \tau_0$,  
 where $T$ and $p$ exhibit 
  a small  peak  of order $2 \times 
 10^{-4}T_c$ and $2\times 10^{-3}p_c$, respectively, 
 with a duration time  about $50\tau_0$. Their 
 detailed behavior can be seen in the insets of Fig.14.  
 At long times, 
$T$ tends to $0.9216T_c$,  
but $p$ continues  to increase slowly.

In Fig.15,  the  profiles of    
$n(r,z,t)$  vs $z$ are given 
 at $r/H=0.1$ at three times 
 together with their cross-sectional  profiles.  
The film thickness decreases gradually   around  
$t=1000\tau_0$ and is stationary at  $t=2000\tau_0$. 
  Figure  16  gives  the time evolution 
of the surface density $n_s(r,t)$ 
at $r/L=0.1$   and 
the  excess  desorption $\Delta \Gamma(t)$ defined in Eq.(3.16). 
Here,  $n_s(r,t)$ increases abruptly around $t=1000\tau_0$ as in 
Fig.11.  However, the decrease in  
 $\Delta\Gamma(t)$ is steep in the early stage 
 $t \ls  200 \tau_0$ and is gradual later  
 until $t \sim 1.3 \times 10^3 \tau_0$.  
Evaporation from   a  thick film is significant  
in the early stage  before 
the transition. 
 
In Fig.17, we show $T(r,z,t)$ in gradation 
at $t/\tau_0=500, 1280$, and $1500$ around  the film. 
The film location is  indicated by 
 a line on which $|\nabla n|$ is largest along $\nabla n$.    
The velocity  $\bi v$ is  displayed   
by arrows.  We can see    a  downward flow 
from the liquid region to the   film induced by the 
film shrinkage.  
The reference  arrow below each panel represents 
the maximum velocity, being equal to 0.284, 3.54, 
and 0.0215 at $t/\tau_0=500, 1280$, and $1500$, 
respectively,  in units of $10^{-2}\ell/\tau_0$.
The shrinking  velocity of the film is of the same order.
In contrast to the  cooling case in Fig.12, 
the region above the film 
is somewhat  heated even in the early stage (at $t/\tau_0=500$), 
which is   because of the film shrinkage in Fig.11. 
The heating is most enhanced at $t/\tau_0= 1280$ 
in the middle plate  
during the pre-dewetting transition. 
The profile at $t=1500\tau_0$ is nearly stationary, 
where a balance is attained 
between evaporation   from the side 
and  condensation from the upper surface. 
Finally, we give the heat flux 
$Q_b(r,t)= - \lambda \p T/\p z$  at $z=0$ 
from the wall to the fluid  at $r=0$ and $L/2$. That is,   
$(-Q_b(0, t), -Q_b(L/2,t)= 
(1.24, 1.22)$ at $t=500$, 
$(44.2, 0.870)$ at $t=1280$, and 
$(1.02, 0.871)$ at $t=1500$ 
in units of $10^{-4} \epsilon\ell / v_0\tau_0$. 
The heat flux is much enhanced  at the center 
during the pre-dewetting transition. 
In the inset of the upper panel of Fig.13, 
we can see that  $\p T/\p z$ is large 
at $r=0$ and is  much smaller 
at $r=L/2$.

\begin{figure}[htbp]
\begin{center}
\includegraphics[scale=1.0]{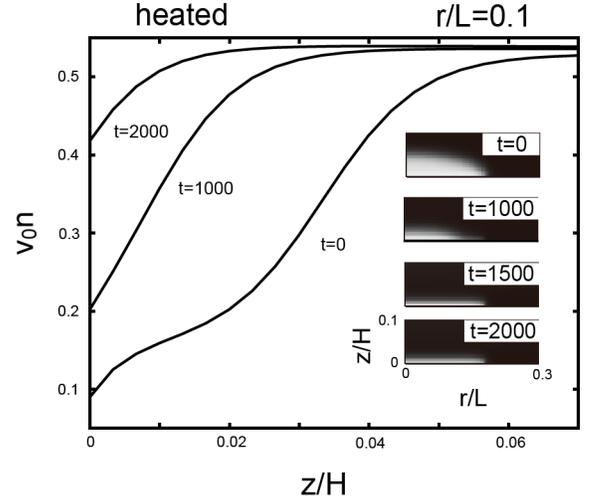}
\caption{\protect
$v_0 n(r,z,t)$
vs $z/H$  at $r/L=0.1$ 
for  $t/\tau_0=0$, $1000$, and $2000$  from below after 
 heating the top.  
The film thickness  decreases as in Fig.16. 
It is stationary  at $t=2000\tau_0$. 
Cross-sectional film profiles are also given 
in the right, where the darkness represents 
the density at four times.  
}
\end{center}
\end{figure}

\begin{figure}[htbp]
\begin{center}
\includegraphics[scale=1.0]{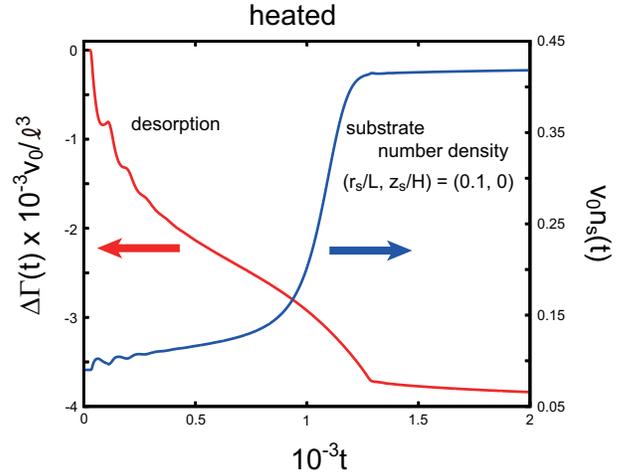}
\caption{\protect
(Color online) 
Normalized excess desorption  $v_0\ell^{-3}
\Delta \Gamma(t)$ in Eq.(3.16) 
 and surface density  $v_0n_s(r,t)$ vs $t$   
 at $ r/L= 0.1$  in units of $\tau_0$ 
 after  heating  the top.    
Here $n_s$ changes abruptly during 
the pre-dewetting transition in the time region 
$800 \tau_0 \ls t\ls 1300\tau_0$, but 
$\Delta \Gamma(t)$ decreases steeply 
in the initial stage $t \ls 200 \tau_0$.  
The system becomes stationary   for 
$t \gs 1300\tau_0$.  
}
\end{center}
\end{figure}

\begin{figure}[htbp]
\begin{center}
\includegraphics[scale=1.0]{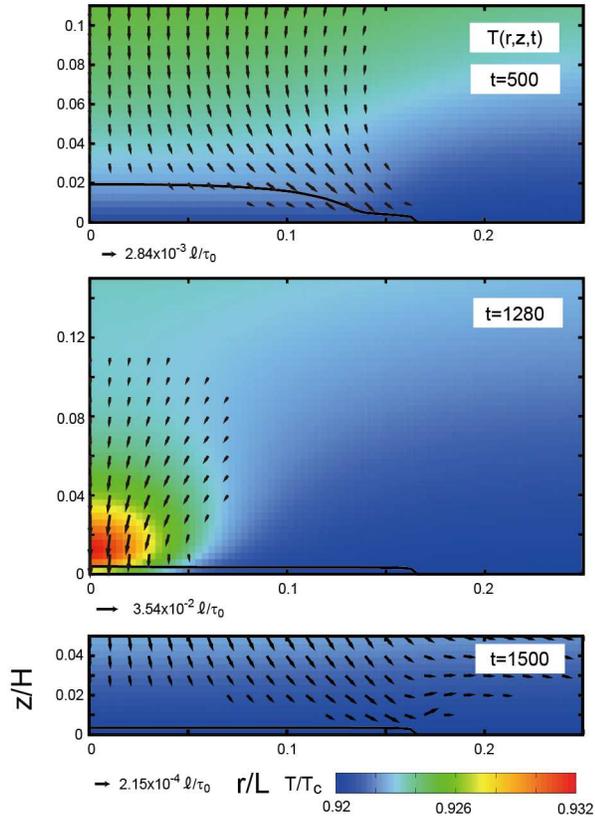}
\caption{\protect
(Color online) 
Temperature $T(r,z,t)$ at $t/\tau_0=
500, 1280$, and $1500$ around the 
hydrophobic spot. On  the block line, 
$|\nabla n|$ is largest. 
Arrows indicate the velocity  $\bi v$. 
Their  amplitudes are  known  from reference 
arrows  below the  panels. 
Even the early stage, the film thickness 
decreases and the region above the film 
is heated due to latent heat release.  
The profile at $t=1500\tau_0$ is nearly stationary. 
}
\end{center}
\end{figure}

\section{Summary and remarks}

We summarize our main results.\\
(i) In Sec.II, for one-component fluids,  we 
 have  examined  the  pre-dewetting transition 
on a  hydrophobic wall   
in the mean-field theory. 
We have assumed the Ginzburg-Landau free 
energy for the number density $n$  in the bulk 
and the surface free energy  
linear in $n$ with $\gamma $ 
representing the  interaction 
between the wall and the fluid. 
Depending on the sign of $\gamma$,  
a line  of  prewetting or  pre-dewetting appears  
in the gas or liquid  side of the coexistence curve. 
If $\gamma$ is small, the line 
is close to the critical point  as in Eqs.(2.21) 
and (2.22). In numerical analysis in Figs. 3,5, and 6,
we have set $\gamma=0.05 \epsilon\ell$ (where 
$\epsilon$ is defined in eq.(2.2) 
and $\ell$  in Eq.(2.12)).  For water 
a rough estimation yields such an order of magnitude of $\gamma$. 
For real systems it is desirable to 
measure $\gamma$. In Subsec.IID, we have numerically 
realized a localized film on a hydrophobic spot 
 in Eq.(2.26), where a thin 
film is  above a pre-dewetting line at 
$T=0.94T_c$ and a thick one is below it at $T=0.92T_c$. 
\\
(ii) In Sec.III, we have investigated the time 
evolution of a localized film on the hydrophobic spot 
 in the axisymmetric geometry.
Starting with 
the equilibrium films  in Fig.7, 
we have numerically integrated 
Eqs.(3.2)-(3.4) 
after a change of the top  temperature 
from $0.94T_c$ to $0.92T_c$ 
in Subsec.IIIC and from $0.92T_c$ to $0.94T_c$ 
in Subsec.IIID. Cooling at the top 
leads to a  pressure decrease, while heating 
at the top leads to a  pressure increase.
We  have found 
singular behavior at the pre-dewetting transition 
in the hydrodynamic variables 
and the excess 
desorption $\Delta \Gamma (t)$ in Eq.(3.16). 
Most markedly,  the liquid 
above the  film is cooled for 
decompression and is heated for 
compression due to latent heat convective transport from 
the growing or shrinking film, as shown in Figs.8, 12, 13, 
and 17.  A small pressure pulse is emitted from the film 
to propagate through the cell, as shown in Figs.9 and 14. 
This pulse could be detected  experimentally \cite{Miura}.  
\\

We give some  remarks.\\  
(1) A  pre-dewetting line 
as well as a prewetting line 
readily follows near the critical point 
for a small surface field $\gamma$. 
It is then of great interest where we can find a  
pre-dewetting line for water on a given hydrophobic wall 
in the phase diagram.\\
(2) 
In future work, we should examine 
 the  role of the long-ranged van der Waals interaction 
 \cite{PG,Bonnreview,Ebner,Evans} 
 on the pre-dewetting phase transition.\\  
(3)  For water-like polar fluids, 
 a small amount of   impurities 
can  strongly promote phase separation   
due to the solvation effect \cite{Current}, 
though we have treated  one-component fluids only. 
As a solute, we may add a noncondensable 
 gas (such as  CO$_2$),  
hydrophobic particles, or ions in water. 
Such impurities 
can strongly affect 
 the formation of  surface bubbles or films 
on a hydrophobic wall in water 
\cite{Higashi,Zhang,Attard,Loshe}.
We will report shortly on 
how the pre-dewetting line is shifted downward 
with increasing  the gas concentration. 
\\ 
(4) 
If the mesh length $\Delta x=\ell/2$ 
is a few $\rm \AA$, our system length  is on the order of 
several ten manometers and the particle number treated 
is of order $10^7-10^8$ (see the beginning of Sec.III). 
Our continuum 
description should be imprecise on the angstrom scale.  
Thus examination  of our results by very large-scale 
molecular dynamics simulations should  be informative. 
We should also investigate  
how our numerical results can be used or modified  
for much larger  film  sizes.\\  
(5) 
Phase changes  inevitably  induce 
a  velocity field carrying heat and mass.
It has been crucial during the pre-dewetting 
transition. We have found 
a steady flow  around the   film at long times 
in Figs.12 and 17. 
In our previous simulation \cite{Teshi2}, 
a steady  circular liquid 
film  was  realized on a homogeneous  wall  
in the complete wetting condition 
under a critical  heating rate, where 
evaporation  and 
condensation  balanced. 
It evaporated to vanish for stronger heating, 
while it expanded for weaker heating or for cooling. 
\\ 
(6) We should study the two-phase 
hydrodynamics such as evaporation or boiling, where 
the hetrogeneity of the wall is  crucial, 
as exemplified in this paper.  
In fluid mixtures, 
a Marangoni flow decisively governs 
the dynamics with increasing the droplet or bubble size 
even at very small solute concentrations 
\cite{Maran}.  \\ 

\vspace{-5mm}

\begin{acknowledgments}
This work was supported by 
the Global COE program 
``The Next Generation of Physics, Spun from Universality and Emergence" 
of Kyoto University 
 from the Ministry of Education, 
Culture, Sports, Science and Technology of Japan. 
R. T. was supported by the Japan Society for Promotion of Science.
We would like to thank Dr. Ryuichi Okamoto for informative 
discussions. 
\end{acknowledgments}

\vspace{10mm} 
\noindent{\bf Appendix: Calculations 
near the critical point }\\
\setcounter{equation}{0}
\renewcommand{\theequation}{A\arabic{equation}}

We examine the pre-dewetting 
transition near the critical point 
in the mean-field theory. 
 The surface free energy density is linear in $n$  as in Eq.(2.1). 
 The density $n$ and the temperature $T$ are assumed to be close 
to their critical values.   We  
expand the Helmholtz free energy density $f= f(n,T)$ 
in powers of the density deviation 
$\psi=n-n_c$ up to the quartic  order as  
\be 
f(n.T) = f_c(T) + \mu_c(T) \psi 
+ \frac{a_0}{2}\tau \psi^2+ \frac{u_0}{4} \psi^4,
\en 
where  $f_c(T)=f(n_c,T)$,  $\mu_c(T)=\mu(n_c,T)$, and 
\be 
\tau=T/T_c-1 
\en  
is the reduced temperature. 
We assume $\tau<0$. Then 
the liquid and gas densities on the coexistence 
line are $n_c +\psi_e$ and $n_c-\psi_e$, respectively,  with 
\be 
\psi_e= (a_0|\tau|/u_0)^{1/2}.
\en  
In the  van der Waals model in Eq.(2.2) \cite{Onukibook},  
$T_c= 8\epsilon/27k_B$,  $n_c=v_0^{-1}/3$, 
and   
\be 
a_0=2\epsilon v_0,\quad u_0= 9\epsilon v_0^3/2. 
\en   
With the Landau expansion (A1), the grand potential density 
$\omega$ in Eq.(2.7) is written as 
\be 
\omega = \frac{u_0}{4} (\psi-\psi_0)^2 (\psi^2 +2\psi_0\psi 
+3\psi_0^2 -2\psi_e^2),
\en 
where $\psi_0=n_0-n_c$ is the value of  
$\psi$ far from the wall assumed to be small. 
The derivative $\omega'= 
\p \omega/\p n$ is equal to the chemical 
potential deviation $\mu-\mu_0$ written as 
\be 
\omega'= u_0(\psi-\psi_0) 
(\psi^2 +\psi_0\psi+\psi_0^2-\psi_e^2). 
\en 
Here we  introduce 
 a dimensionless surface field $\hat\gamma$ by  
 \bea 
 \hat{\gamma}&=& \gamma(2/u_0 M)^{1/2}\psi_e^{-2}\nonumber\\
 &=& 9\sqrt{3}\gamma/8\epsilon \ell |\tau|,
 \ena 
where the second line is the result of   the van der 
Waals model. For each $\hat{\gamma}$ 
and $\psi_0/\psi_e$,  
the condition (2.15) yields the equation 
for  the surface density 
deviation $\psi_s=n_s-n_c$ (which is smaller than $\psi_0$)  
in the form, 
\be 
(\psi_0-\psi_s)(\psi_s^2 +2\psi_0\psi_s 
+3\psi_0^2 -2\psi_e^2)^{1/2}/\psi_e^2=  \hat{\gamma}.
\en
As illustrated  in Fig.2, 
 this equation has two solutions 
$\psi_s^L= n_s^L-n_c$ and $\psi_s^H= n_s^H-n_c$, 
corresponding to the two surface densities  
at the pre-dewetting transition.

First, we take 
 the limit $n_0 \to n_{\rm cx}^\ell$ or $\psi_0 \to \psi_e$ 
on the prewetting line,  
where the film thickness  grows 
 as in Eq.(2.20) but $\psi_s^L$, 
 $\psi_s^H$, and $\hat\gamma$ 
 tend to finite  limiting values. 
From   Eq.(A8) we 
obtain $\hat{\gamma}= |\psi_s^2/\psi_e^2-1|$ in this limit.
Since $\psi_s^L<-\psi_e$ and $\psi_s^H>\psi_e$ we find 
\bea 
\psi_s^L=
n_s^L-n_c &=&- \psi_e(1+\hat{\gamma})^{1/2}, \nonumber\\
\psi_s^H=n_s^H-n_c &=& \psi_e(1-\hat{\gamma})^{1/2}. 
\ena    
The equal area requirement  $S_1=S_2$ in Fig.2 yields 
\be 
(1+\hat{\gamma})^{3/2}
-(1-\hat{\gamma})^{3/2}=2, 
\en 
whose square  gives 
$3\hat{\gamma}^2-1= (1-\hat{\gamma}^2)^{3/2}$. 
Again taking the  square of this  equation,  
we find $\hat{\gamma}^4+6{\hat \gamma}^2=3$, 
which is solved to give   
\be 
\hat{\gamma}  = \sqrt{2\sqrt{3}-3}=0.68125. 
\en  
Therefore, as $T\to  T_{\rm cx}^{\rm dw}(\gamma)$, 
the two surface densities at the pre-dewetting transition are 
$\psi_s^L\cong  -1.2966\psi_e$, 
and $\psi_s^H \cong  0.5646\psi_e$.  
From Eq.(A7), the pre-dewetting value of 
$\gamma$ on the coexistence curve is written as  
\be 
\gamma_{\rm cx}(T)
= \frac{8}{9} (2/\sqrt{3}-1)^{1/2} |\tau|\epsilon \ell.
\en 
We then obtain the coefficient $A_{\rm cx}$ 
in Eq.(2.24).

Second, 
we seek  the value of $\gamma$ at the pre-dewetting 
critical point, denoted by $\gamma_c(T)$. 
Since   $\p^2\omega/\p n^2=0$  at 
$n=n_s$,  $n_s$ coincides with 
the  spinodal   density on the gas  branch so that    
\be 
\psi_s= n_s- n_c= - \psi_e/\sqrt{3}.
\en 
Then  from Eq.(2.15) we find 
\be 
\gamma_c(T)= {8}|\tau| \epsilon \ell/9.
\en  
From the additional 
condition  $\omega' =0$ 
 at $\psi=\psi_s$, Eq.(A7) yields 
  the bulk density deviation $\psi_0$  in this case as   
\be 
\psi_0= n_0- n_c= 2\psi_e/\sqrt{3},
\en 
which is larger than $\psi_e$ as it should be the case. 
For each small $\gamma$, 
the reduced temperature $\tau$ and the density 
deviation $n_0-n_c$ are given by 
$\tau_c^{\rm dw}= -9\gamma/8\epsilon \ell$ 
and $4|\tau_c^{\rm dw}|/3\sqrt{3}v_0$, respectively, at  
the pre-dewetting critical point.  
We then obtain the coefficient $A_{c}$ in Eq.(2.24).

Our calculation results are  applicable to 
the prewetting transition near the critical point, 
where  $\gamma$,  $\psi_0$, and $\psi_s$ are 
small negative quantities. 
 To use the above relations,  
 we should set $\psi_e= -  (a_0|\tau|/u_0)^{1/2}$ 
 and replace  $\gamma$  by $|\gamma|$. 
 See the comment below Eq.(2.23) 
 on  $T_{\rm w}$ and $T_c^{\rm pw}$.

We should note that 
Papatzacos \cite{Papa}  obtained 
the wetting angle $\theta_{\rm w}$ in  partial wetting 
 for the linear surface free energy in Eq.(2.1) 
 and the bulk free energy in Eq.(A1) 
(see Ref.\cite{Y} also). 
In our notation, the dewetting angle $\theta_{\rm dw}=
\pi-\theta_{\rm w}$  satisfies  
\be 
(1+\hat{\gamma})^{3/2}
-(1-\hat{\gamma})^{3/2}=2\cos \theta_{\rm dw}, 
\en 
where the fluid is on the coexistence curve.    
The above equation becomes 
$\hat{\gamma}^6+6\hat{\gamma}^4 +3\hat{\gamma}^2
[1-2\cos (2\theta_{\rm dw})] 
= \sin^2(2\theta_{\rm dw})$. 
Setting  $\beta= 3^{-1}\arccos(\sin^2\theta_{\rm dw})$, we 
solve this equation to obtain 
\be 
\hat{\gamma}= 2{\rm sgn}({\pi}/{2}- \theta_{\rm dw}) 
 \sqrt{\cos\beta(1-\cos\beta) },
 \en 
where     
${\rm sgn}(x)$ gives the sign of $x$. 
For $\gamma>0$, 
 we have  the complete dewetting  limit 
 $\theta_{\rm dw} \to 0$ (or $\theta_{\rm w} \to \pi$) 
as $T \to T_{\rm cx}^{\rm dw}= 
T_c(1- A_{\rm cx}\gamma/\epsilon\ell+\cdots)$ from below. 
For $\gamma<0$,  we have  
the  complete wetting  limit
 $\theta_{\rm w} \to 0$ (or $\theta_{\rm dw} \to \pi$) 
 as $T \to T_{\rm w}= 
T_c(1- A_{\rm cx}|\gamma|/\epsilon\ell+\cdots)$ from below.

\end{document}